\documentclass[twocolumn]{aastex63}
\usepackage{amsmath}
\usepackage{amssymb}
\def\code#1{\texttt{#1}}

\definecolor{purple}{rgb}{0.5,0,0.5}
\definecolor{darkgreen}{rgb}{0.1,0.6,0.1}
\definecolor{orange}{rgb}{1,0.6,0}

\newcommand{\unitstyle}[1]{\ensuremath{\mathrm{#1}}}

\newcommand{\Kelvin}{\unitstyle{K}}
\newcommand{\K}{\Kelvin}  

\newcommand{\Msun}{\ensuremath{\unitstyle{M}_\odot}}

\newcommand{\Rsun}{\ensuremath{\unitstyle{R}_{\odot}}}

\submitjournal{ApJ}
\shorttitle{AGN Stars Spin Fast}
\shortauthors{Jermyn et al.}

\begin{document}

\title{Stellar Evolution in the Disks of Active Galactic Nuclei Produces Rapidly Rotating Massive Stars}

\correspondingauthor{Adam S. Jermyn}
\email{adamjermyn@gmail.com}

\author[0000-0001-5048-9973]{Adam S. Jermyn}
\affiliation{Center for Computational Astrophysics, Flatiron Institute, New York, NY 10010, USA}

\author[0000-0001-6157-6722]{Alexander J. Dittmann}
\affiliation{Center for Computational Astrophysics, Flatiron Institute, New York, NY 10010, USA}
\affil{Department of Astronomy and Joint Space-Science Institute, University of Maryland, College Park, MD 20742-2421, USA}

\author[0000-0002-8171-8596]{Matteo Cantiello}
\affiliation{Center for Computational Astrophysics, Flatiron Institute, New York, NY 10010, USA}
\affiliation{Department of Astrophysical Sciences, Princeton University, Princeton, NJ 08544, USA}

\author[0000-0002-3635-5677]{Rosalba Perna}
\affiliation{Center for Computational Astrophysics, Flatiron Institute, New York, NY 10010, USA}
\affiliation{Department of Physics and Astronomy, Stony Brook University, Stony Brook, NY 11794-3800, USA}

\begin{abstract}

Stars can either be formed in or captured by the accretion disks in Active Galactic Nuclei (AGN). These AGN stars are irradiated and subject to extreme levels of accretion, which can turn even low-mass stars into very massive ones ($M > 100 \Msun$) whose evolution may result in the formation of massive compact objects ($M > 10 \Msun$).
Here we explore the spins of these AGN stars and the remnants they leave behind.
We find that AGN stars rapidly spin up via accretion, eventually reaching near-critical rotation rates.
They further maintain near-critical rotation even as they shed their envelopes, become compact, and undergo late stages of burning.
This makes them good candidates to produce high-spin massive black holes, such as the ones seen by LIGO-Virgo in GW190521g, as well as long Gamma Ray Bursts (GRBs) and the associated chemical pollution of the AGN disk.

\end{abstract}


\section{Introduction} \label{sec:intro}

Active Galactic Nuclei (AGN) are believed to be powered by massive accretion disks draining into supermassive black holes (SMBHs)~\citep{1969Natur.223..690L}.
Because of their crucial role in AGN, these accretion disks have been extensively studied since the pioneering work on their structure by \citet{Shakura1973}. 

Recent years have seen a rekindled interest in AGN disks, particularly in light of gravitational wave detections by LIGO and Virgo.  In particular, the observation of black holes (BHs) with masses above the maximum mass allowed by pair instability in massive stars \citep{PhysRevLett.125.101102}, as well as in the lower mass gap \citep{Abbott2020lowM}, finds a natural explanation in the environments of AGN disks, where compact object mergers are enhanced, and neutron stars (NSs) and BHs grow by accretion due to the very high disk gas densities (e.g. \citealt{McKernan2012, Yang2019, Tagawa2020}). 

The presence of stars and compact stellar remnants in AGN disks is not surprising.
Stars can end up in the disks of AGNs via at least two mechanisms: in-situ formation when disks become self-gravitating and unstable to fragmentation (e.g. \citealt{1980SvAL....6..357K,Goodman2003,Dittmann2020}), and capture from the nuclear star cluster surrounding the AGN as a result of momentum and energy loss as the stars interact with the disk (e.g. \citealt{Artymowicz1993,2020ApJ...889...94M,Fabj2020}). 
Once in an AGN disk, stars may evolve and thereby form compact objects such as black holes and neutron stars.

Stars in AGN disks (AGN stars) are believed to evolve quite differently from those in standard galactic environments.
AGN disks are much hotter and denser than the interstellar medium, so AGN stars are subject to very different boundary conditions than normal stars.
The evolution of these stars has been recently studied by~\citet{2020arXiv200903936C}, who found that AGN stars can quickly become very massive
($M>100~\Msun$) due to rapid accretion fueled by the large gas reservoir in the AGN disk, and that chemical mixing plays a critical role.

The study of~\citet{2020arXiv200903936C} did not, however, investigate the role of rotation in these stars and how it is affected by the special environment of an AGN disk.
Rotation plays a very important role in stellar evolution \citep[e.g.][]{Maeder:2009,Langer:2012} and in determining whether a massive star at the end if its life can produce a long gamma-ray burst (long GRB) during core collapse \citep{MacFadyen1999}.
Understanding the rotation of these stars is also crucial to further constrain the origin of the LIGO/Virgo BHs via their inferred spins.

Here we investigate the spin evolution of stars embedded in AGN disks (AGN stars).
In Section~\ref{sec:stars} we briefly describe our model of stellar evolution in AGN disks, including two key improvements introduced by~\citet{Alex} to the treatment of accretion and mass loss from the approach of~\citet{2020arXiv200903936C}.
In Section~\ref{sec:rotation} we present a model of the rotation rates of these stars, with a focus on the angular momentum changes associated with mass gain/loss and on the stochastic nature of accretion in a turbulent medium.
We analytically explore the consequences of this model in Section~\ref{sec:analytics}, and find that at least during the accretion stage many AGN stars ought to spin up to near-critical rotation rates.
These rapid rotators represent most of the parameter space we are able to explore in our models, and these results are confirmed by our numerical simulations of the coupled rotating stellar evolution of AGN stars in Section~\ref{sec:res}.
We then discuss the implications of these rapid rotators for compact objects, gravitational wave observations, and long GRBs in Section~\ref{sec:discuss}.
We conclude with a brief summary of our results in Section~\ref{sec:conclusions}.

\section{Stellar Models} \label{sec:stars}

We model stellar evolution using revision 15140 of the Modules for Experiments in Stellar Astrophysics
\citep[MESA][]{Paxton2011, Paxton2013, Paxton2015, Paxton2018, Paxton2019} software instrument.
Details of the \code{MESA} microphysics and of other software tools used in our experiments are provided in Appendix~\ref{appen:mesa}. 
Because the stars of interest are embedded in an AGN disk we implement modified surface boundary conditions, accretion, and mass loss following~\citet{2020arXiv200903936C}.

As argued by~\citet{Alex}, the most important correction to the Bondi-Hoyle accretion rate used by~\citet{2020arXiv200903936C} comes from the gravitational influence of the supermassive black hole (SMBH).
Near the SMBH tidal forces truncate the radius at which the stellar gravity dominates the gas dynamics, limiting the accretion rate.
A similar situation occurs when giant planets accrete from a protoplanetary disk, and hydrodynamic simulations have shown that prescriptions truncating accretion to begin at the Hill Radius can accurately predict accretion rates in the viscosity regime relevant to AGN disks \citep{{2020MNRAS.498.2054R},{2021ApJ...906...52L}}. 
Accordingly, we take into account tidal effects from the SMBH when calculating accretion onto AGN stars by truncating accretion to the Hill Radius. 

Additionally, when computing mass loss rates we study the effects of reducing the Eddington luminosity $L_{\rm Edd}$ by a factor of $1-\Omega^2/\Omega_{\rm c}^2$, where 
\begin{align}
	\Omega_{\rm c} \equiv \sqrt{\frac{G M_\star}{R_\star}}
\end{align}
is the critical angular velocity for a rigidly-rotating sphere.
This accounts for the fact that the centrifugal acceleration reduces the effective escape velocity of the star and so reduces the radiative acceleration needed to unbind material from the surface \citep{Maeder:2009,Sanyal:2015}.
We refer to models computed using the unmodified Eddington luminosity as $\Gamma$ models and those computed using the rotationally-reduced Eddington luminosity as $\Gamma-\Omega$ models.
We generally believe the $\Gamma-\Omega$ prescription is more physical, though we show models computed with both prescriptions for comparison.

We initialized our runs with a non-rotating zero-age main sequence solar model.
We then relaxed the boundary conditions and accretion rate over approximately $10^7\mathrm{yr}$ from solar-like to those described by~\citet{2020arXiv200903936C}.
Some models fail during this relaxation.
We have endeavoured to minimize the number of such failures, but given the radical adjustment in conditions it is not surprising that some failed, especially at higher AGN densities.

After relaxation we evolved our models until either an age of $10^9\mathrm{yr}$, a core temperature of $3\times 10^9\,\mathrm{K}$, or failure to converge.
We imposed the age limit because AGN disk lifetimes are believed to be of order $1-100\,\mathrm{Myr}$~\citep{10.1093/mnras/stz135,Martini_2001,Haiman_2001}, and because AGN stars which do not complete their evolution in $10^9\mathrm{yr}$ generally follow standard stellar evolution.
The core temperature limit allows us to halt models during Oxygen burning.
Models reaching these core temperatures are expected to undergo core collapse within a timescale $\sim$yr. 

AGN stars rapidly accrete and reach the Eddington luminosity.
Because of this, their interiors have comparable radiation and gas pressures and hence follow $\gamma=4/3$ polytropes.
This means that they are on the edge of stability, and~\citet{2020arXiv200903936C} argued that this means that any of a variety of processes should result in efficient internal mixing.
Following this argument, we implement their phenomenological mixing prescription and further assume that these stars rotate as rigid bodies.
As a result the rotation of an AGN star is specified entirely by its total angular momentum.

\section{Rotation Evolution}\label{sec:rotation}

We assume that AGN stars only change angular momentum through mass loss or accretion.
Following~\citet{2020arXiv200903936C} we allow AGN stars to both accrete ($\dot{M}_{\rm gain}$) and lose mass ($\dot{M}_{\rm loss}$) at the same time, imagining a multidimensional system with simultaneous inflows and radiation-driven outflows.

Because AGN disks are turbulent environments the angular momentum they accrete is best treated as a random variable.
In Appendix~\ref{sec:stochastic} we describe a method for modelling the total angular momentum of the AGN star as a normally-distributed random variable characterized by a mean $\langle J \rangle$ and a variance $\sigma^2_J$.

In practice, however, the correlation timescale of velocities in the AGN disk is of order the dynamical time in the Bondi sphere, which is short compared with the evolutionary timescale of an embedded star.
As a result we always find that $\sigma^2_J / \langle J\rangle^2 \ll 1$, such that the angular momentum is well-characterized by its expectation value.
While our calculations follow the full stochastic model described in Appendix~\ref{sec:stochastic}, in what follows we consider only the mean angular momentum, which we denote simply by $J$.

We neglect torques due to magnetic coupling between the star and its environment, such that the mean angular momentum of the star evolves due to mass loss and accretion as
\begin{align}
    \frac{d J}{dt} &= \dot{M}_{\rm gain}\;  j_{\rm gain, avg} - \dot{M}_{\rm loss}\;  j_{\rm loss, avg},
    \label{eq:mu}
\end{align}
where $\dot{M}_{\rm gain}$ is the accretion rate, $\dot{M}_{\rm loss}$ is the rate of mass loss, $j_{\rm gain, avg}$ is the mean specific angular momentum of the accreting material, and $j_{\rm loss, avg}$ is the mean specific angular momentum of the lost material.
The remainder of this section computes the terms appearing in this equation.

\subsection{Mass Loss}

We assume that mass is lost preferentially at the equator of the star and so carries the equatorial specific angular momentum
\begin{align}
    j_{\rm loss} = \Omega R_{\rm equator}^2.
\end{align}
This is a conservative choice in that it maximizes the lost angular momentum; any other choice would result in a smaller $j_{\rm loss}$ and hence faster-rotating stars. For simplicity we assume that $R_{\rm equator}$ equals the mean radius of the star, so that
\begin{align}
    j_{\rm loss} = \frac{J}{M_\star}\left(\frac{M_\star R_\star^2}{I}\right) = \frac{J}{k M_\star},
\end{align}
where
\begin{align}
    I \equiv \int_0^{M_\star} \frac{2}{3} r^2 dm
\end{align}
is the moment of inertia of the star and
\begin{align}
	k \equiv \frac{I}{M_\star R_\star^2}
\end{align}
is the gyration parameter (i.e. the non-dimensional gyration radius).

\subsection{Mass Accretion}

Next, we must calculate the specific angular momentum of the accreting mass.
There are three relevant length-scales for accretion onto an AGN star, namely the Bondi radius, the Hill radius, and the disk scale height.
These three are generally not independent, so we choose to parameterize the accretion physics in terms of the Bondi and Hill radii.

In a stationary, infinite medium we expect accretion onto a lone star to begin at the Bondi radius
\begin{align}
    R_{\rm Bondi} = \frac{2 G M_\star}{c_{\rm s}^2},
\end{align}
because that is the radius at which the escape velocity of the gravitational potential of the star becomes comparable to the sound speed of the medium~\citep{1952MNRAS.112..195B}.
Here $c_{\rm s}$ is the sound speed in the AGN disk and $M_\star$ is the mass of the star.

When the star orbits a SMBH, however, the star's gravity competes not only with the disk pressure but also with tidal forces from the SMBH.
The radius at which the star's gravity dominates over tides from the SMBH is the Hill radius
\begin{align}
    R_{\rm Hill} = r_{\rm orb} \left(\frac{M_\star}{3 M_{\rm BH}}\right)^{1/3},
\end{align}
where $r_{\rm orb}$ is the orbital radius of the star and $M_{\rm BH}$ is the mass of the central black hole. 
To account for this effect we model the accretion as beginning at the smaller of $R_{\rm Bondi}$ and $R_{\rm Hill}$, which we identify as the accretion radius
\begin{align}
R_{\rm acc} \equiv \min{(R_{\rm Bondi}, R_{\rm Hill})}\,.
\end{align}
\citet{Alex} studied a variety of other effects, including vertical density variations, and found them to be comparatively unimportant.

\begin{figure}
    \includegraphics[width=0.45\textwidth]{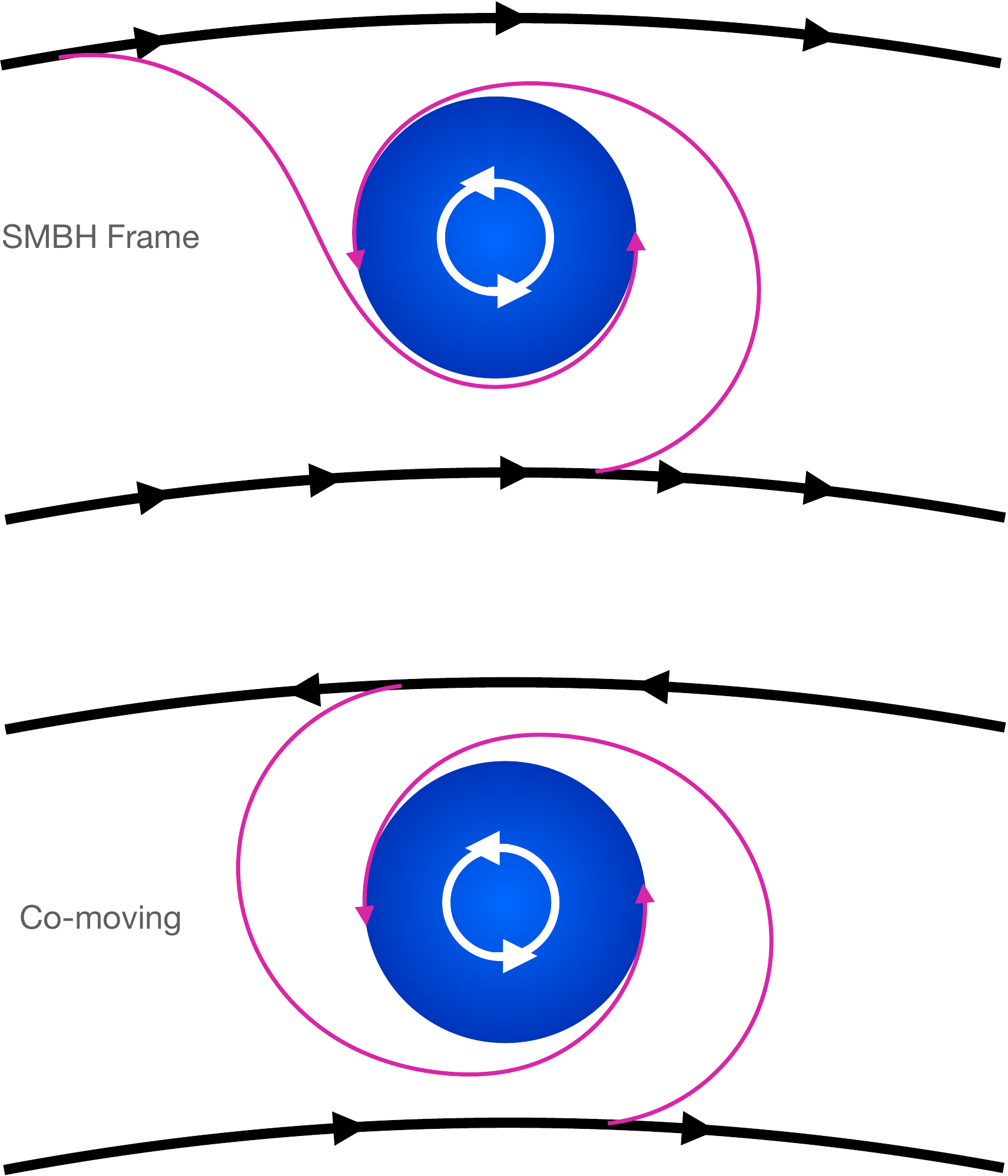}
    \caption{Material accretes from a differentially rotating disk onto a star, spinning the star up. This is shown in the frame of the SMBH (upper) and the frame co-moving with the star (lower).  AGN stars obtain retrograde rotation (backwards with respect to their orbital motion).}
    \label{fig:schema1}
\end{figure}

If there is no viscous or magnetic angular momentum transport across the the accretion radius then the angular momentum of accreting material is just that of the material at the accretion radius in the frame comoving with the star (Figure~\ref{fig:schema1}).
In that frame the average angular momentum within the Bondi radius is given by the differential rotation of the AGN disk, such that
\begin{align}
    j_{\rm gain, avg} \approx  R_{\rm acc}^2 \frac{d(a \Omega_{\rm AGN})}{da} \approx\Omega_{\rm AGN}  R_{\rm acc}^2,
    \label{eq:jgain}
\end{align}
where $a$ is the semi-major axis of the orbit of the AGN star around the central SMBH.
Note that this angular momentum is oriented retrograde relative to the disk because material closer to the SMBH is moving faster than the star, and material further out is moving slower.

So long as
\begin{align}
	j_{\rm gain, avg} < j_{\rm max} = \sqrt{G M_\star R_\star}
\end{align}
material can fall directly from the accretion radius onto the star and it is likely a good approximation to neglect torques at $R_{\rm acc}$.
However, when $j_{\rm gain, avg} > j_{\rm max}$ material must shed angular momentum in order to accrete.
The material likely forms an accretion disk which then transports angular momentum outwards towards and beyond $R_{\rm acc}$ via viscous and/or magnetic torques.
We therefore truncate $j_{\rm gain, avg}$ to be no greater than $j_{\rm max}$.
Implicitly we therefore assume that \emph{excess} angular momentum beyond $j_{\rm max}$ is lost via magnetic/viscous torques across the accretion radius.

\citet{Alex} investigated the effect of modifications to the accretion rate when the accreting angular momentum is large and found that these corrections (1) do not qualitatively change AGN star evolution beyond the effects of tides and (2) are generally less important than the tidal effects which we have included.
We therefore \emph{do not} reduce the accretion rate in this case, though it is likely that the formation of a disk is associated with some reduction in $\dot{M}_{\rm gain}$ and that may have some quantitative effect on the resulting evolution.

\subsection{Truncation}

The star cannot spin faster than its critical rotation rate and so its total angular momentum is limited to
\begin{align}
	J \la J_{\rm crit} \approx k\sqrt{G M_\star^3 R_\star},
	\label{eq:jcrit}
\end{align}
where we have ignored aspherical corrections in this approximation.
To impose the constraint in equation~\eqref{eq:jcrit} we truncate $J$ at the end of each time-step to lie between -$J_{\rm crit}$ and $+J_{\rm crit}$.

\section{Analytic Predictions}\label{sec:analytics}

Before studying our numerical results it is worth deriving some analytical predictions.
For these purposes we focus on AGN stars that undergo runaway accretion, and divide their evolution into three phases:
\begin{enumerate}
	\item Initial runaway accretion.
	\item Constant mass (the ``Immortal Phase'' of~\citealt{2020arXiv200903936C}).
	\item Late-stage super-Eddington mass loss.
\end{enumerate}

\subsection{Runaway Accretion}\label{sec:runaway}

During runaway accretion the accretion rate is always much greater than mass loss.
Moreover, because the star begins with $J=0$ the mean angular momentum is increasing.
Therefore, with $j_{\rm loss,\, avg} \propto \Omega_\star$ we see that $j_{\rm gain,\,avg}$ is typically larger than $j_{\rm loss,\, avg}$.
With these two considerations we neglect angular momentum loss and assume that
\begin{align}
\frac{d  J }{dt} &= \dot{M}_{\rm gain}\, j_{\rm gain,\, avg}\,.
\end{align}
With some rearranging, we find
\begin{align}
\frac{d\ln  J / J_{\rm crit}}{d\ln M_\star} &= \frac{j_{\rm gain,\, avg} M_\star}{J} - \frac{d\ln J_{\rm crit}}{d\ln M_\star}\,.
\end{align}
Inserting equation~\eqref{eq:jcrit} for $J_{\rm crit}$ and treating $k$ as a constant we find
\begin{align}
\frac{d\ln  J / J_{\rm crit}}{d\ln M_\star} &= \frac{j_{\rm gain,\, avg} M_\star}{J} - \frac{3}{2} - \frac{1}{2}\frac{d\ln R_\star}{d\ln M_\star},
\label{eq:dlnJ_Jc}
\end{align}
where the total derivative on the right-hand side is equivalent to $(M_\star/\dot{M})d\ln R_\star/dt$.

We now estimate $d\ln R_\star/d\ln M_\star$.
\citet{2020arXiv200903936C} found that when AGN stars become massive they become radiation-dominated and thus approach $\gamma=4/3$ polytropes.
For a fixed composition this implies that they have approximately fixed gyration parameter $k$\footnote{$k$ is only approximately fixed because the precise definition of $R_\star$ matters, and this is not necessarily a constant as our boundary conditions do depend on $M_\star$. Hence $k$ does vary slightly as the star accretes, though empirically (Section~\ref{sec:res}) we do find that $k$ is approximately constant during this phase of evolution.}.
Because of their high accretion rates they continue to undergo core hydrogen burning throughout the accretion phase.
The rate of nuclear burning is an extremely strong function of core temperature, so this regulates their core temperatures to a narrow window around $\log T/\mathrm{K} \approx 7.5$.
Because AGN stars in this phase are radiation-dominated, we know that
\begin{align}
	P_{\rm c} \approx P_{\rm rad} = \frac{1}{3} a T_{\rm c}^4,
	\label{eq:pc_rad}
\end{align}
where the subscript `c' denotes a quantity evaluated in the core and $a$ is the radiation gas constant.
For a polytrope we also know from scaling considerations that
\begin{align}
	P_{\rm c} \approx \frac{G M_\star^2}{R_\star^4}.
	\label{eq:pc_polytrope}
\end{align}
Combining equations~\eqref{eq:pc_rad} and~\eqref{eq:pc_polytrope} we find
\begin{align}
	R_\star \propto M_\star^{1/2} T_{\rm c}^{-1},
\end{align}
so if $T_{\rm c}$ is approximately constant then
\begin{align}
	R_\star \propto M_\star^{1/2}.
	\label{eq:rscale}
\end{align}

Inserting equation~\eqref{eq:rscale} into equation~\eqref{eq:dlnJ_Jc} we find
\begin{align}
\frac{d\ln  J / J_{\rm crit}}{d\ln M_\star} &= \frac{j_{\rm gain,\, avg} M_\star}{J} - \frac{7}{4}\,.
\end{align}
Expanding $j_{\rm gain,\, avg}$ then yields
\begin{align}
\frac{d\ln  J / J_{\rm crit}}{d\ln M_\star} &= \min\left(\frac{M_\star \Omega_{\rm AGN} R_{\rm acc}^2}{ J},\frac{J_{\rm crit}}{k  J}\right) - \frac{7}{4}\,.
\label{eq:Jgain_evol}
\end{align}
With increasing mass we see that this reaches a fixed point when
\begin{align}
	 \frac{J}{J_{\rm crit}}  = \frac{4}{7}\min\left(\frac{M_\star \Omega_{\rm AGN} R_{\rm acc}^2}{J_{\rm crit}},\frac{1}{k}\right),
\end{align}
that is, the specific angular momentum approaches the smaller of either $4/7k$ times critical or the specific angular momentum of the accreting material.

For many choices of AGN disk parameters, $M_\star \Omega_{\rm AGN} R_{\rm acc}^2 \gg J_{\rm crit}/k$, which means that
\begin{align}
\frac{d\ln  J / J_{\rm crit}}{d\ln M_\star} &= \frac{J_{\rm crit}}{k  J} - \frac{7}{4}.
\end{align}
Hence with increasing mass we see that $J/J_{\rm crit}$ increases rapidly, asymptoting to the fixed point where $ J = (4 /7k)J_{\rm crit}$.
For a sphere of uniform density, $k = 2/5$.
AGN stars have higher densities in their cores than their envelopes and so have $k < 2/5$.
As a result the fixed point has $J = (10/7) J_{\rm crit} > J_{\rm crit}$ and so in our models $J$ grows to $J_{\rm crit}$ and truncates there.

\subsection{Constant Mass}

In the constant-mass regime, accretion is balanced by mass loss.
The mean angular momentum evolves according to
\begin{align}
\frac{d  J }{dt} &= \dot{M}_{\rm gain} j_{\rm gain,\, avg} - \dot{M}_{\rm loss}  j_{\rm loss}\\
	&= \dot{M}_{\rm gain}\left(j_{\rm gain,\, avg} - \frac{ J}{k M_\star}\right).
\end{align}
This has a fixed point when
\begin{align}
	\frac{ J}{k M_\star} = j_{\rm gain,\, avg},
\end{align}
so stars in this evolutionary phase evolve to a specific angular momentum which is proportional to that which they accrete.

When $M_\star \Omega_{\rm AGN} R_{\rm acc}^2 \gg \sqrt{G M_\star R_\star} = J_{\rm crit}/k$ we truncate $j_{\rm gain,\, avg}$ to $\sqrt{G M_\star R_\star} =  J_{\rm crit} / k M_\star$, in which case the fixed point has
\begin{align}
 J  = J_{\rm crit}
\end{align}
and stars tend towards critical rotation.
For $M_\star \Omega_{\rm AGN} R_{\rm acc}^2 \ll J_{\rm crit}/k$, AGN stars will evolve towards sub-critical rotation.


\subsection{Late-Stage Mass Loss}\label{sec:loss}

During late-stage mass loss the rate of mass loss is always much greater than the accretion rate, so the angular momentum evolves according to
\begin{align}
\frac{d  J }{dt} &= \dot{M}_{\rm loss}  j_{\rm loss}  =   J  \frac{\dot{M}_{\rm loss}}{k M_\star}.
\end{align}
Following the same reasoning as in Section~\ref{sec:runaway} and again assuming constant $k$ we obtain
\begin{align}
\frac{d \ln  J / J_{\rm crit} }{d\ln M} &= \frac{1}{k} - \frac{3}{2} + \frac{1}{2} \frac{d\ln R_\star}{d\ln M_\star} \approx \frac{1}{k} - \frac{7}{4}.
\label{eq:Jloss_evol}
\end{align}
Because density increases towards the center of the star we have $k < 2/5$, which means that $k < 4/7$ as well, so $ J/J_{\rm crit}$ becomes smaller as the star loses mass.
Note, however, the factor of $J_{\rm crit}/ J$ difference between equation~\eqref{eq:Jgain_evol} and~\eqref{eq:Jloss_evol}: during the accretion phase the star evolves towards critical faster than it loses angular momentum in the mass loss phase.
As a result we expect stars to net gain angular momentum during their evolution.

In particular, so long as $M_\star \Omega_{\rm AGN} R_{\rm acc}^2 \gg J_{\rm crit}/k$ we expect that most AGN stars eventually reach critical rotation.
When they subsequently lose mass they spin down.
How far they fall below critical depends on how much mass they lose and on the gyration parameter $k$, which we cannot estimate in an analytic fashion.

\section{Results} \label{sec:res}

We now turn to the results of our numerical simulations.

\subsection{Time Evolution}

\begin{figure*}
	\includegraphics[width=\textwidth]{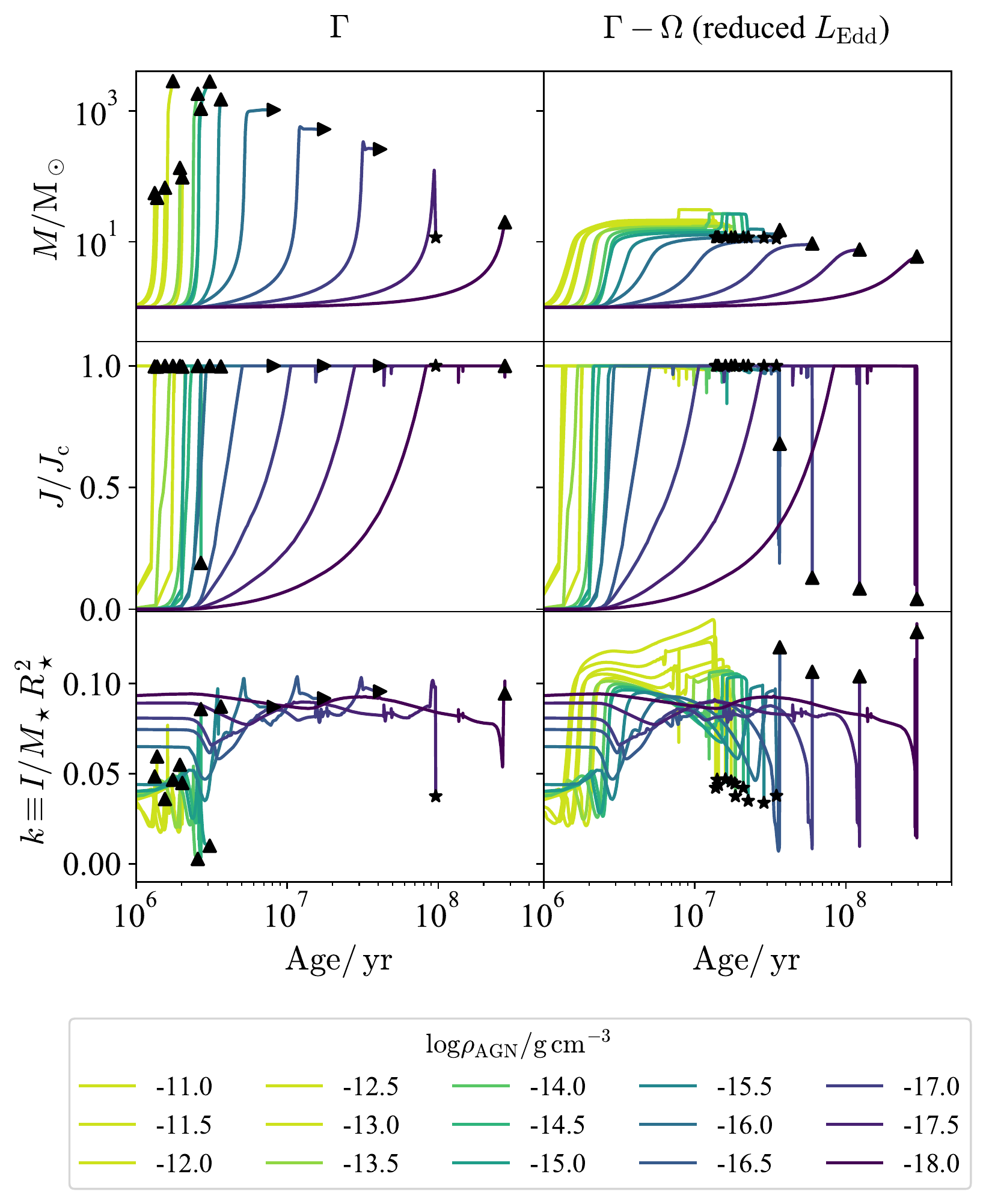}
	\caption{The mass (upper), $ J/J_{\rm c}$ (middle), and gyration parameter $k$ (lower) are shown as functions of stellar age for $\Gamma$ (left) and $\Gamma-\Omega$ (right) models as functions of the AGN density $\rho_{\rm AGN}$. The end of each evolutionary track is labelled by either a star (likely core collapse), an upward arrow (ongoing accretion), or a right-going arrow (immortal phase). Recall that the $\Gamma-\Omega$ models have $L_{\rm Edd}$ reduced to account for rotational effects, and so undergo more rapid mass loss. These models have Keplerian angular velocity $\Omega_{\rm AGN}=10^{-11}\mathrm{rad\,s^{-1}}$ and AGN disk sound speed $c_{\rm s} = 10\,\mathrm{km\,s^{-1}}$. Age is referenced to the end of the atmospheric boundary condition blend period, near the start of accretion.}
	\label{fig:MJK}
\end{figure*}

Figure~\ref{fig:MJK} shows the evolution of mass ($M_\star$, upper), angular velocity relative to critical ($ J /J_{\rm c}$, middle), and gyration parameter ($k$, lower) as functions of time for stellar models with both the $\Gamma$ (left) and rotation-reduced (right, $\Gamma-\Omega$) prescription for the Eddington luminosity and associated mass loss. 
Evolutionary tracks are coloured by AGN density $\rho_{\rm AGN}$, which we swept in the range $10^{-18}-10^{-11}\mathrm{g\,cm^{-3}}$.
We prescribe the AGN temperature by specifying a disk sound speed $c_{\rm s} = 10\,\mathrm{km\,s^{-1}}$, which is the default used by~\citet{2020arXiv200903936C}.
The Keplerian angular velocity $\Omega_{\rm AGN}$ determines the distance from the SMBH and so the strength of tidal effects, and was set to $10^{-11}\mathrm{rad\,s^{-1}}$.

For both the $\Gamma$ and $\Gamma-\Omega$ $L_{\rm Edd}$ prescriptions we see that models above an AGN disk density of $10^{-18}\mathrm{g\,cm^{-3}}$ exhibit rapid runaway accretion, in agreement with the findings of~\citet{2020arXiv200903936C} and \citet{Alex}.
In the case of the $\Gamma$ prescription, models with density above $3\times 10^{-18}\mathrm{g\,cm^{-3}}$ are then immortal, exhibiting a balance between accretion and mass loss which replenishes fresh hydrogen in their cores.
Models below this critical density eventually deplete their core hydrogen and evolve, rapidly losing mass to become high-metallicity compact objects.
In the case of the $\Gamma-\Omega$ prescription, mass loss generally wins over accretion in the end, and we see most models evolve into compact objects.

In each case the initial accretion causes the stars to spin up to critical rotation.
Surprisingly, the stars which lose mass remain critically rotating.
This is not what we predicted in Section~\ref{sec:loss}, and the reason for the discrepancy is that there we assumed that the gyration parameter is constant throughout the star's evolution, whereas we see from the lower row of Figure~\ref{fig:MJK} that stars become much more compact as they enter later stages of nuclear burning, resulting in a lower gyration parameter.
Thus, even though their specific angular momentum falls, they become compact even faster and remain critically rotating.

The few models which do not end their lives as critical rotators are those which remained at relatively low masses. These models are not chemically or quasi-chemically homogeneous, and proceed to evolve onto the red giant branch. The resulting large radial extension implies these objects rotate with very slow, sub-critical rotation velocities. 
We expect such evolution to dominate at low densities of $\la 10^{-18}\mathrm{g\,cm^{-3}}$ for the $\Gamma$ prescription and $\la 10^{-16}\mathrm{g\,cm^{-3}}$ for the $\Gamma-\Omega$ prescription.

\subsection{Tidal Forces}

\begin{figure*}
	\includegraphics[width=\textwidth]{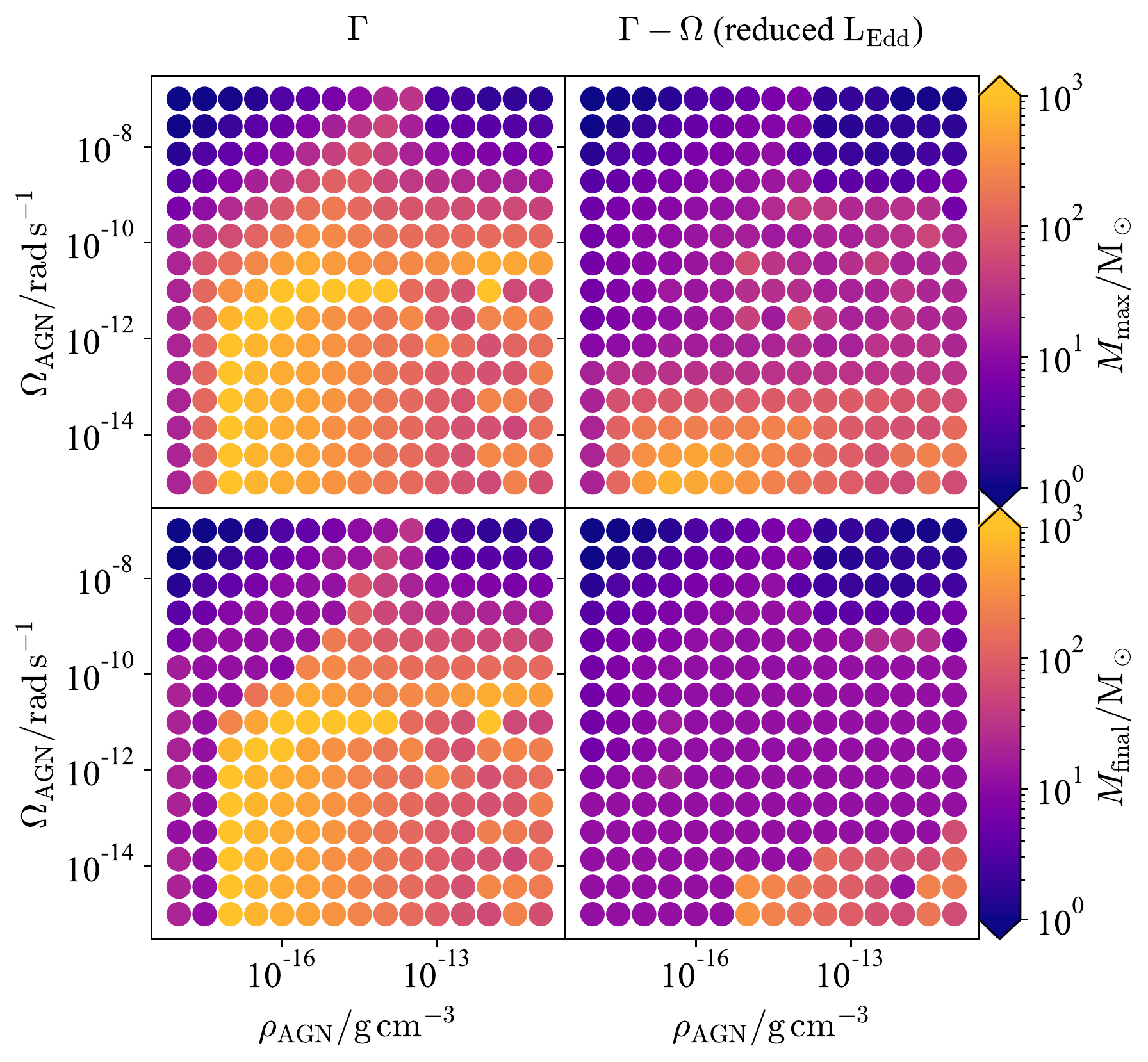}
	\caption{The peak mass (upper) and final mass (lower) in $\Msun$ is shown for each model with the $\Gamma$ (left) and $\Gamma-\Omega$ (right) prescriptions as functions of the AGN density $\rho_{\rm AGN}$ and Keplerian angular velocity $\Omega_{\rm AGN}$.}
	\label{fig:mass}
\end{figure*}


We next study the role of tides, which we choose to parameterize by $\Omega_{\rm AGN}$.
Figure~\ref{fig:mass} shows the peak mass (upper) and final mass (lower) for models in a grid running over $\rho_{\rm AGN}\in [10^{-18} ... 10^{-11}]\mathrm{g\,cm^{-3}}$ and $\Omega_{\rm AGN}\in [10^{-15} ... 10^{-7}]\mathrm{rad\,s^{-1}}$. 
The range of angular velocities corresponds to radial coordinates in the disk on the order of $10^{3...8}$ gravitational radii, and the range of densities corresponds to of order $10^{3...6}$ gravitational radii.
Here the gravitational radius is
\begin{align}
	r_g = \frac{2G M_{\rm BH}}{c^2},
\end{align}
where $c$ is the speed of light.

As before, we study models with both the $\Gamma$ and $\Gamma-\Omega$ prescription for $L_{\rm Edd}$, which governs the rate of mass loss.
As a reminder the $\Gamma-\Omega$ prescription has a rotationally-reduced $L_{\rm Edd}$ and so undergoes \emph{rotationally enhanced} mass loss.

With increasing $\Omega_{\rm AGN}$ the Hill radius falls below the Bondi radius and accretion onto the star becomes tidally limited.
Thus we see a trend towards decreasing peak and final mass with increasing $\Omega_{\rm AGN}$.
For the same reason more models successfully ran at higher $\Omega_{\rm AGN}$ because they accreted more slowly during the initial model relaxation.

Comparing the $\Gamma$ and $\Gamma-\Omega$ prescriptions we see lower masses in the models where rotation reduces $L_{\rm Edd}$.
This is because as $L_{\rm Edd}$ decreases the threshold for forming a super-Eddington wind decreases, so significant mass loss begins at lower masses than in the regular-$L_{\rm Edd}$ models.

\begin{figure*}
	\includegraphics[width=\textwidth]{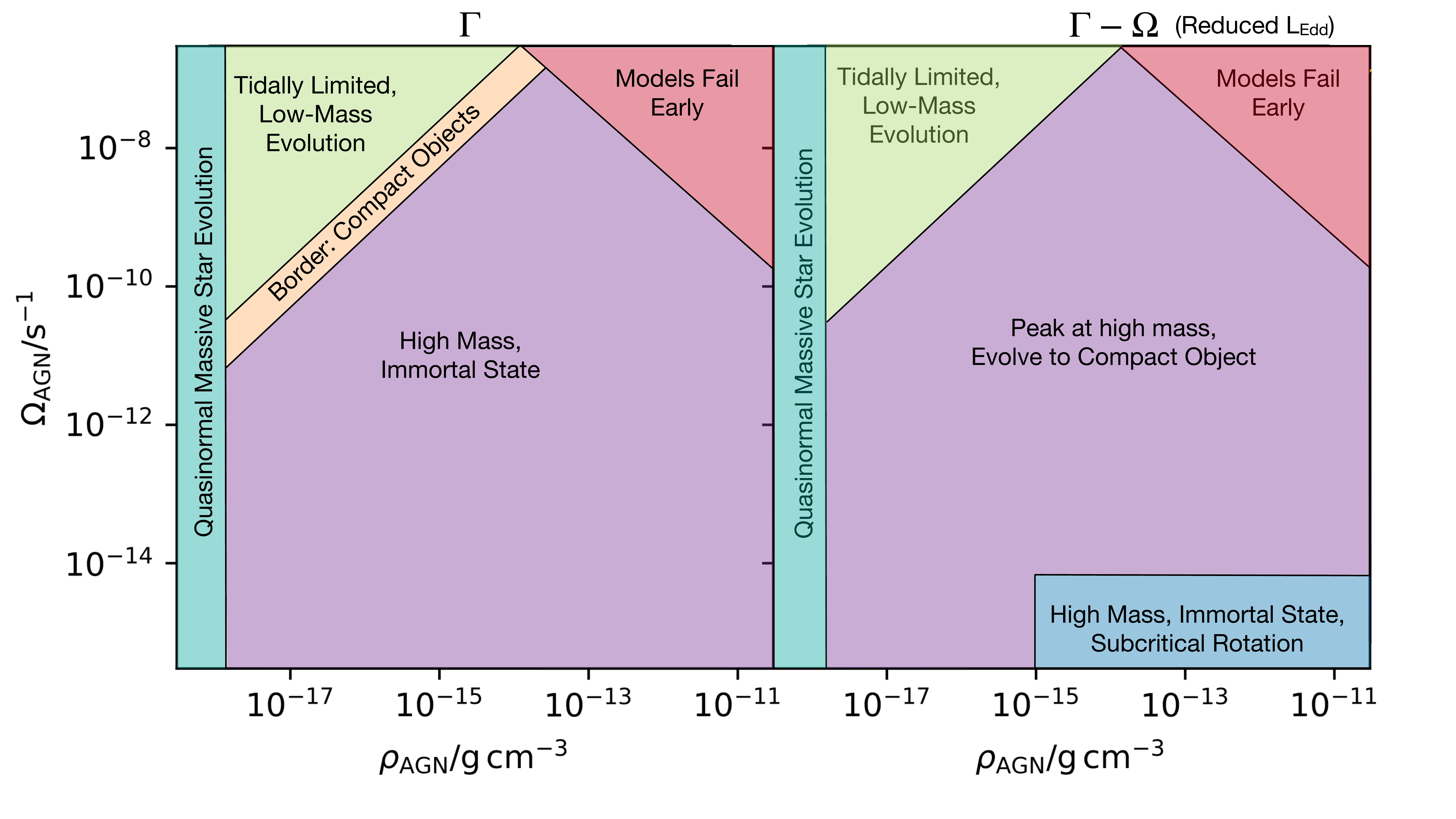}
	\caption{The boundaries between different kinds of evolution discussed in the text are shown for the grids with the $\Gamma$ (left) and $\Gamma-\Omega$ (right) prescriptions as functions of the AGN density $\rho_{\rm AGN}$ and Keplerian angular velocity $\Omega_{\rm AGN}$.}
	\label{fig:schema2}
\end{figure*}

Consistent with~\citet{Alex} in the $\Gamma$ grid we see several different classes of evolution:
\begin{itemize}
	\item At low densities of $10^{-18}\mathrm{g\,cm^{-3}}$ or less stars do not accrete beyond $10 \Msun$ and so follow ordinary massive stellar evolution, going up the giant branch and reaching $R > 100 \Rsun$ (blue region in Figure~\ref{fig:schema2}).
	\item At higher densities and $\Omega_{\rm AGN}^{4/3} / \rho_{\rm AGN} \la 10^{2}\mathrm{cm^3 g^{-1} s^{-4/3}}$ tidal effects are unimportant. Stars rapidly gain mass and reach $10^2-10^3 \Msun$. At that point either the models remain at that high mass in the immortal state or the \code{MESA} models fail to converge. In the latter case we believe that if they converged they would enter the immortal state at an even higher mass (purple region in left panel).
	\item At higher $\Omega_{\rm AGN}^{4/3} / \rho_{\rm AGN} \ga 10^{2}\mathrm{cm^3 g^{-1} s^{-4/3}}$ the Hill radius falls below the Bondi radius for a wide range of stellar masses. This limits the accretion rate and thereby limits the peak mass. However, these objects do not undergo ordinary stellar evolution, because they are near the Eddington limit and rapidly exchange material with the AGN disk via accretion and mass loss (green region).
	\item Models at very high $\rho_{\rm AGN}$ and $\Omega_{\rm AGN}$ accrete rapidly but \code{MESA} fails to converge too early into their evolution for us to even guess as to what will happen at late times (red region).
\end{itemize}
In addition to the above, a small fraction of models on the boundary between regions (2) and (3) peak in mass around $100 \Msun$, burn through all of their Hydrogen, and undergo rapid mass loss down to $10 \Msun$, eventually becoming compact helium/carbon/oxygen stars.
A more detailed characterization of this behaviour is provided by~\citet{Alex}.

Because many stars become rapid rotators, when we reduce $L_{\rm Edd}$ to account for rotational effects ($\Gamma-\Omega$) we see the boundaries between these classes shift considerably:
\begin{itemize}
\item Rather than an unusual edge-case, a majority of $\Gamma-\Omega$ models with $\Omega_{\rm AGN}^{4/3} / \rho_{\rm AGN} \la 10^{2}\mathrm{cm^3 g^{-1} s^{-4/3}}$ peaks around $30-1,000 \Msun$. This accretion spins them to near-critical rotation and enhances mass loss. They then deplete in hydrogen and rapidly shrink to become compact $10 \Msun$ helium/carbon/oxygen stars (purple region on right in Figure~\ref{fig:schema2}).
\item At $\rho \ga 10^{-15}\mathrm{g\,cm^{-3}}$ and $\Omega_{\rm AGN} \la 10^{-14}\mathrm{rad\,s^{-1}}$ the accreting angular velocity is not enough to spin stars up to critical. The resulting reduction in $L_{\rm Edd}$ is small, so these objects accrete without bound until \code{MESA} fails to converge. We expect these would become immortal if \code{MESA} were able to follow their evolution further region (dark blue region on right in Figure~\ref{fig:schema2}).
\end{itemize}

A schematic summarizing the mass evolution we see for both the $\Gamma$ and $\Gamma-\Omega$ grids is shown in Figure~\ref{fig:schema2}.

\begin{figure*}
	\includegraphics[width=\textwidth]{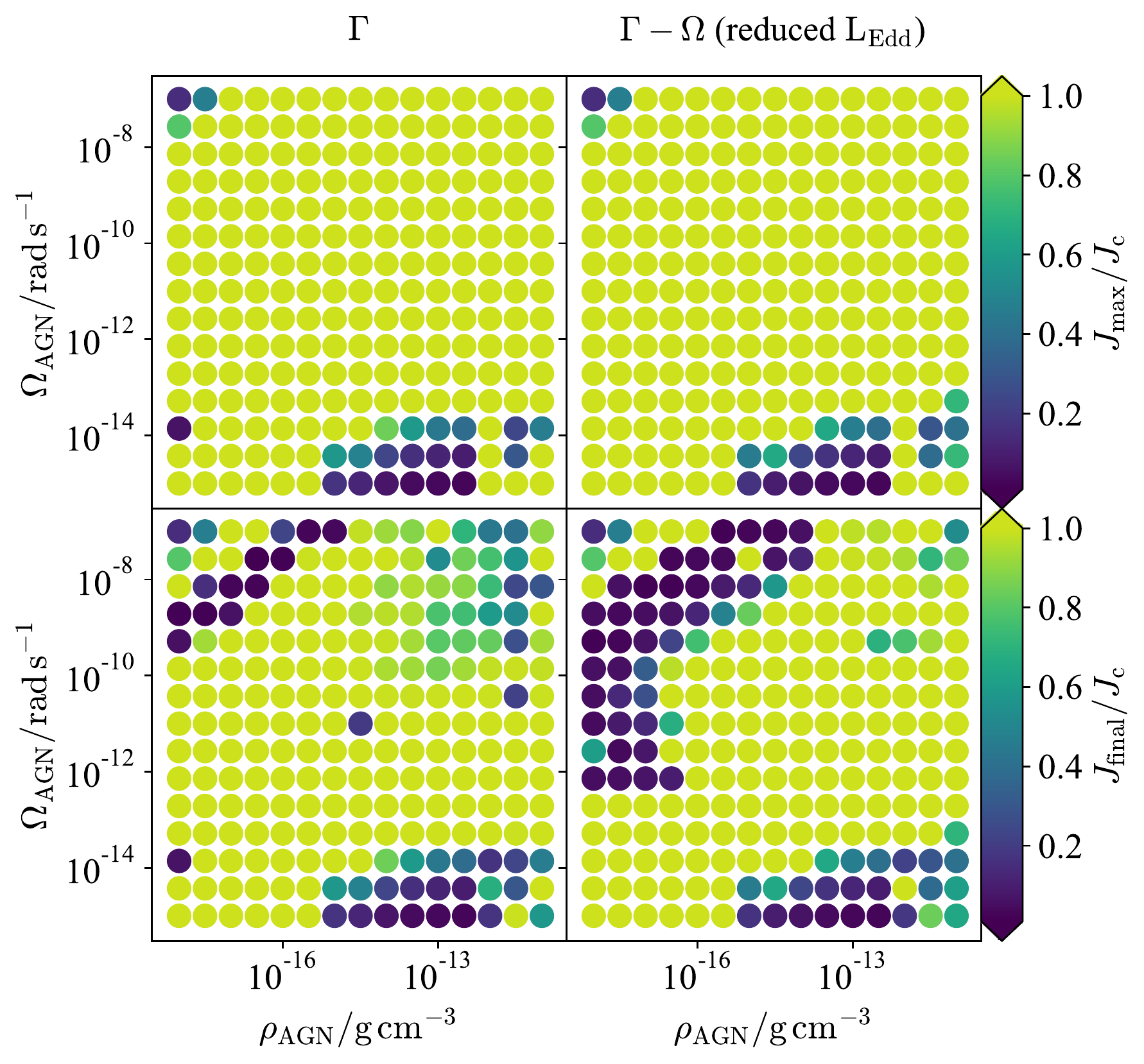}
	\caption{The peak ratio $ J /J_{\rm c}$ (upper) and at the end of the run (lower) is shown for each model with the $\Gamma$ (left) and $\Gamma-\Omega$ (right) prescriptions as functions of the AGN density $\rho_{\rm AGN}$ and Keplerian angular velocity $\Omega_{\rm AGN}$.}
	\label{fig:J}
\end{figure*}

Figure~\ref{fig:J} shows the corresponding angular momentum evolution.
The upper row shows the critical ratio $ J /J_{\rm c}$ at the time of peak mass and the lower row shows the same at the end of the run.
In nearly all cases the models reach critical rotation by the time their masses peak and remain critical through the end of the run.
The models which do not reach critical rotation at any point are almost all at very low $\Omega_{\rm AGN}$ and high $\rho_{\rm AGN}$.
These accrete quickly, but the infalling material has low specific angular momentum and so does not spin them up to critical.
\code{MESA} then fails to converge when the runs reach very large masses, so we never see if these models are immortal or become compact objects.
Interestingly models at the same $\Omega_{\rm AGN}$ but lower density do become critical, because while they accrete to a sub-critical rotation rate they subsequently become compact stars which ultimately makes them critical.

A number of models reach critical rotation at some point but end their evolution sub-critical.
In the $\Gamma$ grid these follow the line $\Omega_{\rm AGN}^{4/3}/\rho_{\rm AGN}\approx 10^{2}\mathrm{cm^{3}\,g^{-1}}\,s^{-4/3}$.
Models in this population accrete slowly because tidal effects truncate $R_{\rm acc}$ to be less than $R_{\rm Bondi}$.
This slow accretion means that models only make it up to a mass of roughly $10 \Msun$, at which point they undergo relatively normal stellar evolution and run up the Red Giant Branch (RGB).
This causes them to inflate substantially, raising $J_{\rm c}$ and thereby lowering $J/J_{\rm c}$.

In the $\Gamma-\Omega$ grid the same thing happens but over a wider range of parameter space, extending down to $\Omega_{\rm AGN} \approx 10^{-12}\mathrm{rad\,s^{-1}}$ at $\rho_{\rm AGN} \la 10^{-17}\mathrm{g\,cm^{-3}}$.
More models end up on the RGB in this grid because they experienced rotationally-enhanced mass loss.

In both grids, models at higher $\Omega_{\rm AGN}$ than this population accrete slower and so end up with $M_\star \approx 2 \Msun$ after $10^9\mathrm{yr}$.
Because the specific angular momentum of the infalling material is high and $k \la 1/2$ this is enough to cause them to rotate critically, though they have not yet reached their peak mass.

\begin{figure*}
	\includegraphics[width=\textwidth]{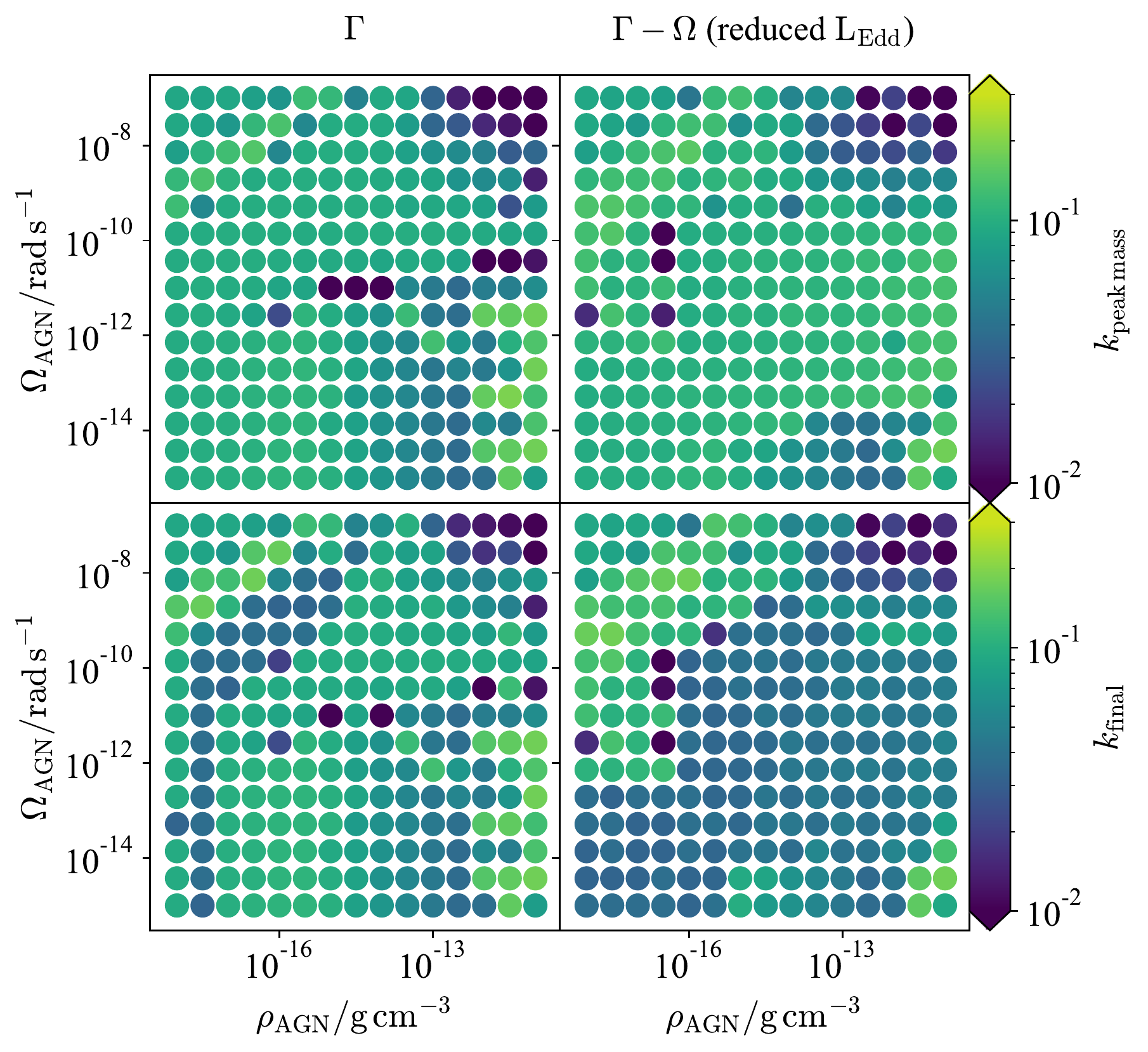}
	\caption{The gyration parameter $k = I/M_\star R_\star^2 $ at the time of peak mass (upper) and the end of the run (lower) is shown for each model with the $\Gamma$ (left) and $\Gamma-\Omega$ (right) prescriptions as functions of the AGN density $\rho_{\rm AGN}$ and Keplerian angular velocity $\Omega_{\rm AGN}$.}
	\label{fig:k}
\end{figure*}

The majority of our models accrete mass and angular momentum, reach critical rotation, and are then either immortal or remain critically rotating through the subsequent mass loss.
Those which undergo mass loss remain critical for the same reason as the models in Figure~\ref{fig:MJK}: as models lose mass they also proceed to later stages of nuclear burning and become more compact.
This is shown in Figure~\ref{fig:k}, which plots the gyration parameter $k$ at peak mass (upper) and the end of the run (lower).
Comparing with Figure~\ref{fig:mass} we see that models which lose substantial amounts of mass are much more compact at the end of their mass loss than at their peak mass (i.e. $k$ falls with time).
This is particularly evident in the runs with $\Gamma-\Omega$ prescription, most of which shed 90\% of their mass from peak to end.

In both grids the models which do not end in a critical state are split into two populations.
There is a cluster at high densities and low $\Omega_{\rm AGN}$ for which \code{MESA} fails to converge before they reach their maximum masses.
However, we believe by analogy with models at slightly lower densities which \emph{do} converge that these models would become critical if we were able to follow their evolution through into the compact object state.
The second population are at low densities and high $\Omega_{\rm AGN}$.
These stars accrete slowly enough that they become giants rather than reaching extreme masses and then forming compact objects.
These giant models have enormous moments of inertia and so even though they spend most of their lives with near-critical rotation they die as slow rotators.

\section{Astrophysical Implications}\label{sec:discuss}

To summarize our findings, in regions of the AGN disk where $\rho_{\rm AGN} > \rho_{\rm crit} \approx 10^{-18}\mathrm{g\,cm^{-3}}$ we expect captured stars to undergo rapid accretion.
This density scale increases towards the SMBH as tidal effects become more important, and decreases with sound speed as $\rho_{\rm crit}\propto c_s^{-3}$~\citep{Alex}.

Because there is an angular velocity gradient in the AGN disk, the disk has a net vorticity in the frame of a co-orbiting star.
On scales of the Bondi radius this vorticity means that gas in the disk has a large, typically super-Keplerian, specific angular momentum in the frame of the star.
When that gas accretes, even if limited to the Keplerian angular velocity, it serves to rapidly spin the star up to critical even very far out in the disk where $\Omega_{\rm AGN}$ is as low as $3\times10^{-14}\mathrm{rad\,s^{-1}}$.

Depending on the exact physical prescriptions used, many of these stars then evolve through later stages of nuclear burning, undergo rapid mass loss, and become compact $10 \Msun$ high-metallicity objects.
We expect core collapse to occur soon after, though we have not tried to follow the collapse process in \code{MESA}~\citep{2020arXiv200903936C}.
Because these stars become more compact as they lose mass they remain critical rotators through the end despite shedding a large fraction of their peak angular momentum.
They are therefore good candidates for producing long GRBs and fast-spinning black holes.

The stars which do not lose mass enter an immortal phase~\citep{2020arXiv200903936C} where fresh hydrogen-rich material is accreted fast enough to continuously replenish the core.
This phase ends when these stars either enter a low-density pocket of the AGN disk or when the disk itself dissipates.
In either case mass loss then comes to overwhelm accretion and the immortal phase ends, with stars rapidly evolving towards the same compact $10 \Msun$ state as before~\citep{2020arXiv200903936C}.
We expect this to again end in core collapse of a critically-rotating star.

We now examine the prospects for producing long GRBs and rapidly rotating black holes.

\subsection{Production of GRBs, and their observability in AGN disks} \label{sec:GRB}

\begin{figure}
	\centering
	\includegraphics[width=0.48\textwidth]{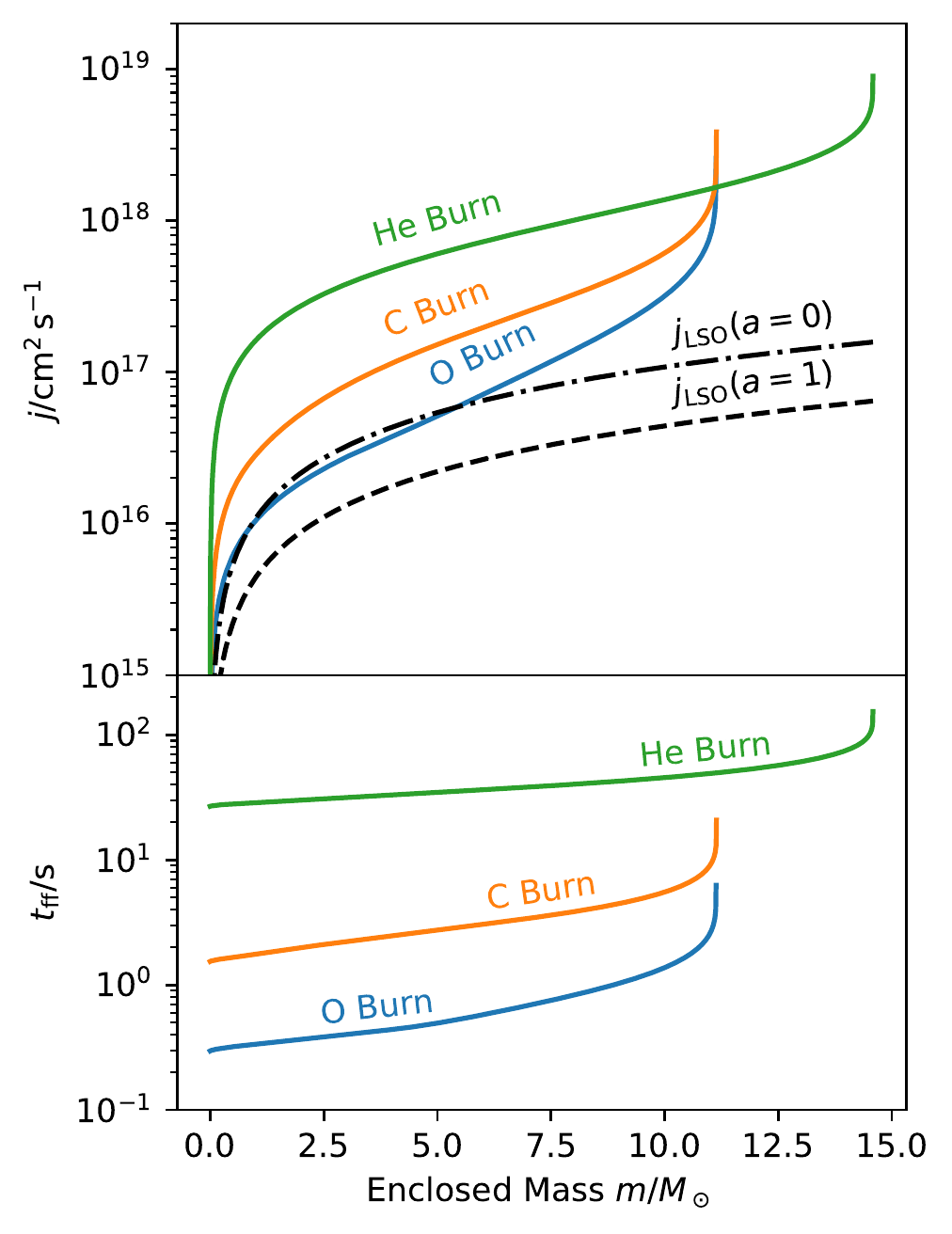}
	\caption{(Upper) The specific angular momentum profile in an AGN star model is shown as a function of enclosed mass coordinate $m$. Also shown are the specific angular momenta of least-stable orbits of a black hole of mass $m$ and spins $a=0$ and $a=1$. (Lower) The free-fall time  is shown as a function of enclosed mass. Three models are shown from a single evolutionary track computed with the $\Gamma-\Omega$ prescription, $\rho_{\rm AGN}=10^{-16}\mathrm{g\,cm^{-3}}$, and $\Omega_{\rm AGN}=10^{-11}\mathrm{rad\,s^{-1}}$. The models were chosen at the onset of helium burning, the onset of carbon burning, and late stages of oxygen burning.}
	\label{fig:isco}
\end{figure}

Our results, and in particular the fact that 
massive stars in AGNs are found to be fast rotators, bear important implications for 
long GRBs. These are found to be a fraction of
$\sim 0.5-4\%$ of SNe-Ibc in the local Universe (e.g. \citealt{Dellavalle2006}), and are known to be associated with very energetic supernovae from the  collapse of massive  stars \citep{Hjorth2003,Stanek2003}, as suggested by theoretical models \citep{MacFadyen1999,Woosley2006,Yoon2006}.
The $\gamma$-ray emission is believed to be produced within a relativistic jet (e.g. \citealt{Piran1999}), and a key element to launch a jet is believed to be an hyperaccreting disk around a BH\footnote{Note however some models assume a magnetar engine \citep{Thompson:2004,Metger2011}.} \citep{MacFadyen2001}. 

For an accretion disk to be formed, 
a fraction of gas must remain bound after the supernova explosion, and it must posses a specific angular momentum $j_m$ at least as large as the specific angular momentum of the last stable orbit, $j_{\rm lso}$.  
Our results (cfr. upper panel of 
Fig.~\ref{fig:isco}) show that close to core collapse the envelopes of AGN stars are endowed with enough angular momentum to produce an accretion disk around a newly formed BH ($j_m>j_{\rm lso}$ virtually for any BH with non-zero spin). Hence we draw the important conclusion that, upon their death, {\em massive stars in AGN disks are typically expected to produce long GRBs.} It is interesting to compare these stars to the Wolf-Rayet  progenitors of standard long GRBs from field stars \citep{MacFadyen1999,Yoon2006,Woosley2006,Cantiello:2007}. The similarity is not surprising, since here we assume that AGN stars accreting large amounts of mass are well mixed \citep{2020arXiv200903936C}. Hence they evolve quasi-chemically homogeneously, similarly to rapidly rotating long GRB progenitors.

The amount of mass which is available for accretion is given by the fraction which remains bound after the supernova explosion (examples are given in Fig.6 of \citealt{Perna2014}). For weak explosions, most of the material falls back.

The material that remains bound falls back on a timescale which is on the order of the free-fall time~\citep{2012ApJ...752...32W}
\begin{equation}                        
t_{\rm ff} (r)= \frac{1}{\sqrt{24G\bar{\rho}}}\,,     
\label{eq:tff}                      
\end{equation}
where $\bar{\rho}$ is the mean density of the star.
 The precise distribution of the initial fallback radii for all the bound particles will clearly depend on the details of the supernova explosion, but a minimum
value is given by the location of the particles prior to the explosion.
The free-fall time of the envelope (bottom panel of Fig.~\ref{fig:isco}) is on the order of a few tens of seconds, as in the bulk of the typical range of durations of  long GRBs. 

After the bound material falls back, it circularizes at a radius $R_{\rm circ}$ determined by the condition $j(R_{\rm circ}) = j_{\rm m}$.  Subsequently, the evolution of the disk is determined by the viscous timescale 
\begin{equation}                                t_0\,(R_{\rm circ})\,=\,\frac{R_{\rm circ}^2}{H^2 \alpha \Omega_K}\,                               
\sim 0.9\; \alpha^{-1}_{-1}\;m^{-1/2}_{3}\,    
R_{8}^{3/2}\left(\frac{R}{H}\right)^{2}\,{\rm s} \;,                                              
\label{eq:tvisc}
\end{equation}
where $m_{3}=M/(3~\Msun)$, $R_{8}=R/(10^{8}~{\rm cm})$, $\Omega_K$ is the Keplerian
velocity of the gas in the disk, $H$ the disk scale-height, and
$\alpha$ the viscosity parameter, written in
units of $\alpha_{-1}\equiv\alpha/0.1$ \citep{Shakura1973}.
At early times, while fallback still goes on, the accretion rate is determined by the longest between $t_{\rm ff}$ and $t_0({\rm R_{circ}})$. In the inner parts of the disk, up to hundreds of gravitational radii, the scale is set by the free-fall one, yielding accretion rates on the order of $\dot{m}_{\rm acc}\sim 0.01-0.1 \Msun$~s$^{-1}$ over several tens of seconds, as typical of long GRBs. 

With massive stars in AGN disks possessing the key elements to power a long GRB upon their death, the next question to address is the likelihood to observe such sources as they emerge from the dense environments of AGN disks. The question of the observability of relativistic, electromagnetic transients produced in AGN disks
was recently addressed by \citet{Perna2021}, considering two specific models for the disk structure, the one by \citet{Sirko2003} (SG in the following) and the one by \citet{Thompson2005} (TQM in the following). The location of the sources was assumed to be in the disk's mid-plane, which is the most pessimistic case in terms of observability. This turns out to be also the most likely occurrence, since it is expected that most of the stars interacting with the disk should end up in its mid-plane (e.g. \citealt{Tanaka2004}). 

The analysis by \citet{Perna2021} showed that the outcome is quite dependent on the disk model and on the SMBH mass, which relates to disk properties, such as the density and the radial extent. Long GRBs occurring in
disks around SMBHs of mass $\sim 10^6 \Msun$, are expected to appear as typical transients (that is similar to the ones occurring in standard galactic environments) for most locations of the disk, except for some regions, $\sim 10^4$-a few $\times 10^5 R_g$ ($R_g$ being the gravitational radius) in the SG disk, and between a few $\times 10^5-10^6 R_g$ in the TQM disk model, when the prompt emission and early afterglow emerge on a timescale set by the diffusion time
\begin{equation}
    t_{\rm diff} \approx \frac{H^2 \rho_0 \sigma_T}{m_p c},
\label{eq:tdiff}
\end{equation}
where $H$ is the scale height of the disk and $\rho_0$ the density in the disk mid-plane, $\sigma_{\rm T}$ the Thomson cross section, and $m_p$ the proton mass.

For AGNs with SMBHs of larger masses, the regions in which both the prompt GRB emission and the afterglow appear normal are gradually reduced: the increasing opacity of the disk causes the transients to be diluted on the diffusion timescale. The magnitude of this timescale varies from minutes to several years, being generally smaller in the inner disk regions and for less massive AGN disks \citep[see Fig.5 in][for quantitative details]{Perna2021}.

\subsection{Black Hole Spins} \label{sec:BH}


Our models predict that stars embedded in AGN disks of densities $\rho_{\rm AGN} > 10^{-18}\mathrm{g\,cm^{-3}}$ typically grow to large masses ($M_{\rm max} > 50 \Msun$).
The exceptions are those very near the SMBH with $\Omega_{\rm AGN} > 10^{-11}\mathrm{rad\,s^{-1}}$, where tidal effects slow accretion.
Even these less massive stars, however, typically evolve into massive helium/carbon/oxygen stars of roughly $10 \Msun$ (see lower row of Figure~\ref{fig:mass}) which, owing to their compactness, likely undergo core collapse and form black holes~\citep{2020arXiv200903936C}.

Our models also suggest that AGN stars end their lives with near-critical rotation.
This then seems to be the typical fate of AGN stars, to form black holes with mass $M_{\rm BH} \approx 10 \Msun$ and spin $a \approx 1$.

The resulting black holes may then go on to accrete and form a population of more massive objects~\citep{1952MNRAS.112..195B,1964ApJ...140..796S}.
The same differential rotation in the AGN disk that caused their progenitors to spin up will then maintain their rapid rotation even as they gain mass.
The result is a population of first-generation black holes with $M \ga 10 \Msun$ and $a \approx 1$.
This channel of black hole formation is of particular interest given the recent observation of GW190521, a merger of two mass gap black holes ($M_1 \approx 85 \Msun, M_2 \approx 66 \Msun$) consistent with large spins ($a_1 \approx 0.69^{+0.27}_{-0.62}$, $a_2 \approx 0.73^{+0.24}_{-0.64}$)~\citep{PhysRevLett.125.101102}, though see also~\citet{2021ApJ...907L...9N}.
While we do not have estimates of the formation rates of such objects, these black holes 
could be the descendants of low-mass stars embedded in an AGN disk. After accreting and becoming massive rapid-rotators, they lost part of their mass and formed $\approx10 \Msun$ black holes. These black holes further accreted up from the disk\footnote{At Eddington-limited accretion rates this process would take of order $100-300\mathrm{Myr}$ though, which is longer than we expect AGN disks to live.
If accretion is slightly super-Eddington, by a factor of a few, that would suffice to bring this time-scale down to plausible disk lifetimes.}, or they migrated and merged with other compact remnants, reaching their pre-GW190521 masses \citep[e.g.][]{McKernan2012,Secunda:2019,Yang2019,Tagawa2020}. 

A challenge to this picture is that the observed merger was most consistent with spins misaligned with the orbital plane~\citep{PhysRevLett.125.101102}, while a strong prediction of our models is that the black hole spins should be aligned with the rotation of the AGN disk.
While it is certainly possible for the orbital plane of the merging binary to be misaligned with the disk, this does not seem to be the most likely scenario, which suggests that other physics we have not considered could be at work in GW190521.

\section{Summary}\label{sec:conclusions}

Stars in AGN disks are thought to evolve in a wide variety of unusual ways.
Depending on the AGN disk density and sound speed and the strength of tidal forces we expect AGN stars to
\begin{itemize}
\item Cease to age,
\item Accrete up to masses over $10^3 \Msun$,
\item Undergo quasi-chemically homogeneous evolution,
\item Shed the vast majority of their mass,
\item Spin up to critical rotation, and/or
\item Form compact helium/carbon/oxygen stars.
\end{itemize}
All of these behaviours are exhibited by stellar models that began with the same zero-age main-sequence $1 \Msun$ initial condition, and the variation we see is entirely a function of the conditions in the AGN disk.

Here we studied the spin evolution of AGN stars, as well as its impact on the accretion and evolution of these objects.
At densities $\rho_{\rm AGN} \ga 10^{-18}\mathrm{g\,cm^{-3}}$ AGN stars accrete rapidly.
When that gas accretes it serves to rapidly spin AGN stars up to critical, even very far out in the disk where $\Omega_{\rm AGN}$ is as low as $3\times10^{-14}\mathrm{s^{-1}}$.
We find that rotational enhancement of mass loss is then important, causing most of these stars to then rapidly lose mass and evolve into compact, critically-rotating $10 \Msun$ objects made of helium and heavier elements.

We expect most of these compact stars to eventually undergo core collapse, generating Gamma Ray Bursts and leaving behind rapidly-spinning black holes.

For stars very near the central supermassive black hole we expect accretion to be tidally limited, resulting in slightly less massive, longer-lived stars which nonetheless reach critical rotation.
These could potentially form a population of rapidly-rotating stars which persist for several million years after the AGN disk dissipates.

\acknowledgments

The Flatiron Institute is supported by the Simons Foundation. RP acknowledges support by NSF awards AST-1616157 and AST-2006839 and from NASA (Fermi) award 80NSSC20K1570.
We thank Yuri Levin for insightful comments on the accretion of angular momentum.

\software{
\texttt{MESA} \citep[][\url{http://mesa.sourceforge.net}]{Paxton2011,Paxton2013,Paxton2015,Paxton2018,Paxton2019},
\texttt{MESASDK} 20190830 \citep{mesasdk_linux,mesasdk_macos},
\texttt{matplotlib} \citep{hunter_2007_aa}, 
\texttt{NumPy} \citep{der_walt_2011_aa},
\texttt{RemoteExperiments} \url{https://github.com/adamjermyn/remote_experiments}
         }

\clearpage

\appendix

\section{Software Details} \label{appen:mesa}

We performed calculations using revision 15140 of the Modules for Experiments in Stellar Astrophysics
\citep[MESA][]{Paxton2011, Paxton2013, Paxton2015, Paxton2018, Paxton2019} software instrument.

The MESA EOS is a blend of the OPAL \citep{Rogers2002}, SCVH
\citep{Saumon1995}, FreeEOS \citep{Irwin2004}, HELM \citep{Timmes2000},
and PC \citep{Potekhin2010} EOSes.

Radiative opacities are primarily from OPAL \citep{Iglesias1993,
Iglesias1996}, with low-temperature data from \citet{Ferguson2005}
and the high-temperature, Compton-scattering dominated regime by
\citet{Buchler1976}.  Electron conduction opacities are from
\citet{Cassisi2007}.

Nuclear reaction rates are from JINA REACLIB \citep{Cyburt2010} plus
additional tabulated weak reaction rates \citet{Fuller1985, Oda1994,
Langanke2000}.  (For MESA versions before 11701): Screening is
included via the prescriptions of \citet{Salpeter1954, Dewitt1973,
Alastuey1978, Itoh1979}. (For MESA versions 11701 or later):
Screening is included via the prescription of \citet{Chugunov2007}.
Thermal neutrino loss rates are from \citet{Itoh1996}.

We performed extensive convergence testing on both the $\Gamma$ and $\Gamma-\Omega$ model grids using more than 1,600 models to study the dependence of our results on time-step and mesh resolution.
We determined that our results are independent of resolution for spatial resolution parameter \code{mesh\_delta\_coeff} up to 1 and time resolution parameter \code{time\_delta\_coeff} up to 0.2.
Note that this time resolution requires the use of our custom time-step controls.

The final configuration files and code used in our model grids are available in~\citet{adam_s_jermyn_2021_4562499}.
These are given in \code{Python}~3 Pickle files which specify the changes to make to the configuration on top of a base configuration given in the file `inlist\_project'.
These Pickle files also provide the short-sha's of \code{git} commits which can be found in the git repository stored in~\citet{adam_s_jermyn_2021_4562499}.
Each such commit corresponds to a single \code{MESA} run directory used to perform one of our runs, including the full configuration files and `run\_star\_extras' code used.
We further provide the Pickle files specifying the configurations and short-sha's of commits we used in the final set of convergence tests which demonstrate that our results are converged.

The same \code{git} repository contains a history of nearly all \code{MESA} runs used to develop this work contributed to this work.
These are commits whose messages contain the word `patch' and which do not lie on any branch.
These \code{git} experiments were performed using the \code{RemoteExperiments} software package, details of which may be found at \url{https://github.com/adamjermyn/remote_experiments}.



\section{Stochastic Angular Momentum Evolution}\label{sec:stochastic}

The gas in AGN disks is believed to be turbulent, with a characteristic length-scale of the disk scale height $H$ and characteristic velocity-scale of the sound speed $c_s$.
This turbulence imparts a random additional component to the angular momentum which accretes onto a star embedded in the disk.

To model this system we treat the total angular momentum of the AGN star as a normally-distrubuted random variable with mean $\langle J \rangle$ and variance $\sigma^2_J$.
The mean evolves according to the differential equation
\begin{align}
    \frac{d\langle J \rangle}{dt} &= \dot{M}_{\rm gain}  j_{\rm gain, avg} - \dot{M}_{\rm loss}  j_{\rm loss, avg},
    \label{eq:mu2}
\end{align}
which just says that the mean angular momentum increases according to the mean accreted angular momentum $j_{\rm gain,avg}$ and decreases according to the mean lost angular momentum $j_{\rm loss,avg}$.
The variance evolves according to a similar equation:
\begin{align}
        \frac{d\sigma^2_J}{dt} &= \frac{\dot{M}_{\rm gain}^2 j_{\rm gain, std}^2}{\tau_{\rm turb}} - \frac{1}{k}\frac{\dot{M}_{\rm loss}}{M_\star}\sigma_J^2.
        \label{eq:var}
\end{align}
We obtain the first term by assuming that the accreted material follows a random walk in $j$ with characteristic time-scale $\tau_{\rm turb}$ and step size $j_{\rm gain, std}$ set by the structure of turbulence in the disk.
We obtain the second term by assuming that the lost material has the specific angular momentum of the surface $j_{\rm loss} = J/k M_\star$, where
\begin{align}
	k \equiv \frac{I}{M_\star R_\star^2}
\end{align}
is the gyration parameter and $I$ is the moment of inertia of the star.

Equation~\eqref{eq:mu2} is the same as equation~\eqref{eq:mu}, just with a different notation to emphasize the fact that $J$ is a random variable.
We evaluated all terms in this equation in Section~\ref{sec:rotation}, so all that remains is to evaluate the terms $\tau_{\rm turb}$ and $j_{\rm gain, std}$ which appear in equation~\eqref{eq:var}.

The angular momentum within the accretion radius varies stochastically due to turbulence in the disk, giving rise to the term $j_{\rm gain, std}$.
This is just the standard deviation of the specific angular momentum within the accretion stream, which we approximate by
\begin{align}
    j_{\rm gain, std} \approx v_{\rm turb}^2 \tau_{\rm turb},
    \label{eq:jstd}
\end{align}
where $v_{\rm turb}$ is the turbulent velocity at the scale of the accretion radius and $\tau_{\rm turb}$ is the characteristic time-scale of the turbulence at that length-scale.
We can estimate $\tau_{\rm turb} \approx R_{\rm acc} / v_{\rm turb}$ and $v_{\rm turb} \approx c_{\rm s} (R_{\rm acc} / H)^n$ for $R_{\rm acc} < H$, where
\begin{align}
    H \approx \frac{c_{\rm s}}{\Omega_{\rm AGN}},
\end{align}
$H$ is the scale height of the disk, $n$ is an index which depends on the nature of the turbulence, and $\Omega_{\rm AGN} \equiv \sqrt{G M_{\rm BH}/a^3}$ is the angular velocity of the orbit of the star.
In the inertial range $n=1/3$ for incompressible Kolmogorov turbulence~\citep{1941DoSSR..30..301K} and $n=2/3$ for compressible Burgers turbulence~\citep{BURGERS1948171,2013MNRAS.436.1245F}.
Because $v_{\rm turb}$ only equals $c_{\rm s}$ on the outermost scale of $H$, we use the Kolmogorov scaling and write
\begin{align}
    v_{\rm turb} \approx c_{\rm s}\min\left[1, \left(\frac{R_{\rm acc}}{H}\right)^{1/3}\right]
\end{align}
and
\begin{align}
    \tau_{\rm turb} \approx \frac{\min(H, R_{\rm acc})}{v_{\rm turb}}.
\end{align}
We can then evaluate equation $j_{\rm gain, std}$ using equation~\eqref{eq:jstd}.

Finally recall that we cannot allow the star to rotate super-critically.
To incorporate this constraint, after each time-step we check if $\langle J \rangle$ is super-critical.
If it is, we truncate it to the nearer of $\pm J_{\rm crit}$ and set $\sigma^2_J=0$.
More sophisticated mappings are possible~\citep[e.g.][]{RePEc:eee:csdana:v:21:y:1996:i:1:p:119-119}, but in testing we found that when our models attain critical rotation the variance $\sigma^2_J$ rapidly diminishes, making this approach a good approximation.

\bibliography{refs}
\bibliographystyle{aasjournal}

\end{document}